\documentclass[prl,twocolumn,floatfix,showpacs]{revtex4}
\usepackage{amsmath,amsfonts,amssymb}
\usepackage[dvips]{graphicx}
\usepackage{ulem}
\newcommand{\nc}{\newcommand}
\nc{\rnc}{\renewcommand}
\nc{\nn}{\nonumber}
\nc{\db}{\displaybreak[0]\\}
\nc{\ds}{\displaystyle}
\rnc{\[}{\left[}
\rnc{\]}{\right]}
\rnc{\(}{\left(}
\rnc{\)}{\right)}
\nc{\lt}{\left\{}
\nc{\rt}{\right\}}
\nc{\om}{\omega}
\nc{\sg}{\sigma}
\nc{\lam}{\lambda}
\rnc{\a}{\alpha}
\nc{\bra}{\langle}
\nc{\ket}{\rangle}
\rnc{\i}{{\rm i}}
\rnc{\d}{{\rm d}}
\nc\lamt{\tilde{\lam}}
\nc\At{\tilde{A}}
\nc\Mt{\tilde{\mathcal{M}}}

\begin{document}
\title{
Relaxation dynamics of the
asymmetric simple exclusion process 
with Langmuir kinetics
on a ring
}
\author{Jun Sato}
\author{Katsuhiro Nishinari}
\affiliation{
Research Center for Advanced Science and Technology, University of Tokyo, \\\it 
4-6-1 Komaba, Meguro-ku, Tokyo 153-8904, Japan
}
\date{\today}
\begin{abstract}
We consider the asymmetric simple exclusion process with Langmuir kinetics 
on a periodic lattice. 
We analytically obtain the exact time evolution of correlation functions with arbitrary length 
starting from the initial state with no particle in the system. 
The exact stationary state of this model has been known for the totally asymmetric case. 
We propose a basis transformation which simplifies the proof of the stationarity of this state 
and enables the generalization to the partially asymmetric case. 
Moreover, we construct low-energy excitations and obtain the exact relaxation time. 
\end{abstract}
\maketitle
{\it Introduction. }
Nonequilibrium statistical mechanics has been intensively studied in recent years. 
How and why a nonequilibrium state reaches thermal equilibrium is a long-standing problem. 
Despite much effort in this field, 
complete understanding of relaxation dynamics starting from a specific initial state 
is far from being fulfilled. 
As for the static property of a nonequilibrium steady state, however, exact solutions are known for some simple models. 

Among them, the asymmetric simple exclusion process (ASEP) is one of the most fundamental and nontrivial model 
describing the nonequilibrium transport phenomena \cite{MGPB, DEHP, KSKS, Sasamoto, Schadschneider, SCN}, 
which have a wide range of applications such as 
biological transport \cite{MGPB}, 
and pedestrian and traffic flows \cite{Schadschneider, SCN}. 
The ASEP is a continuous time Markov process 
describing the asymmetric diffusion of particles with an exclusion principle on a one-dimensional lattice. 
In a time interval $\d t$, 
a particle hops to the right (left) site with the probability $p\d t$ $(q\d t)$ if it is vacant. 

The most outstanding advantage of this model is the amenability to 
exact analytical methods such as matrix product ansatz \cite{BE} and Bethe ansatz \cite{GM}. 
The steady state of the ASEP is exactly constructed by the matrix product ansatz, 
which offers exact derivations of interesting phenomena such as boundary induced phase transitions \cite{BE, BECE, Krug}. 

As for the dynamics, however, knowledge about the full spectrum of the Markov matrix is needed 
for the complete understanding of the relaxation process. 
As a first step toward this goal, 
spectral gaps of the first excited states are examined 
by use of the Bethe ansatz method, which provides asymptotic information about dynamics such as relaxation time. 
The finite size scaling of it reveals that 
the model is governed by the Kardar-Parisi-Zhang universality class \cite{Dhar, GS, Kim, GM, dGE, AKSS}. 

Furthermore, full relaxation dynamics in the totally asymmetric simple exclusion process (TASEP) on a ring is exactly examined 
by numerical implementations of the Bethe ansatz equations and the determinant formula for form factors \cite{MSS}. 
In this paper, we present an analytical expression of the time evolution of the correlation functions 
in the ASEP with Langmuir kinetics (ASEP-LK) model on a periodic lattice. 

The ASEP-LK model describes an attachment and detachment of particles 
as well as the exclusive hopping process \cite{PFF}. 
The most notable point in this model is the nonconservation of the number of particles even in the closed ring. 
Steady state properties of this model with open boundary conditions have been studied in detail 
by use of mean field approximation and Monte Carlo simulations, 
which reveals interesting phenomena such as 
the coexistence of high- and low-density phases separated by the shock wave in the density profile \cite{EJS, MH}. 
The exact stationary state is constructed in the case of TASEP-LK with a periodic lattice \cite{EN}, 
in which particle hopping occurs only in one direction ($q=0$). 

In this paper we introduce a basis transformation which simplifies the proof of the stationarity of this state 
and enables generalization to the partially asymmetric case ($q\neq0$). 
We construct low-energy excitations and obtain the exact relaxation time. 
Furthermore, we analytically obtain the exact time evolution of correlation functions with arbitrary length 
starting from the initial state with no particle in the system. 

\begin{figure}
\includegraphics[width=0.8\columnwidth]{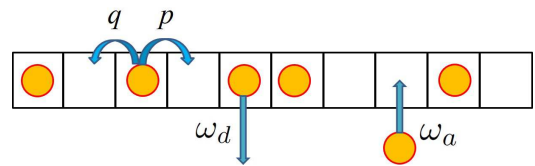}
\caption{
ASEP with Langmuir kinetics
}
\label{model}
\end{figure}

{\it Model. }
We consider the ASEP with Langmuir kinetics described in Fig. \ref{model}. 
A particle is attached on a site with rate $\om_a$ if it is vacant 
and detached from the site with rate $\om_d$ if it is occupied. 

Hereafter we impose a periodic boundary condition with length $L$. 
We associate a Boolean variable $\tau_n$ to every site $n$ 
to represent whether a particle is present $(\tau_n=1)$ or not $(\tau_n=0)$. 
Let $|\tau_n=0\ket$ and $|\tau_n=1\ket$ denote the standard basis vectors in this order for the vector space $\mathbb{C}^2$. 
We consider the $L$-fold tensor product of this basis 
$|\tau_1\tau_2\cdots\tau_L\ket:=|\tau_1\ket\otimes|\tau_2\ket\otimes\cdots\otimes|\tau_L\ket$, 
the dimension of which is $2^L$. 
Then we can write a state of the system at time $t$ in a vector form 
with each element being a probability distribution 
\begin{align}
|P(t)\ket=\sum_{\tau_1,\cdots,\tau_L}
P(\tau_1,\cdots,\tau_L|t)|\tau_1,\cdots,\tau_L\ket. 
\end{align}
The time evolution of this state is described by the master equation
\begin{align}
\frac{\d}{\d t}|P(t)\ket
=\mathcal{M}|P(t)\ket, 
\label{master}
\end{align}
where the Markov matrix $\mathcal{M}$ is given by
\begin{align}
&\mathcal{M}=\sum_{n=1}^{L}\(\mathcal{M}_{n,n+1}+h_n\), 
\nn\\&
\mathcal{M}_{n,n+1}=
\begin{pmatrix}
0& 0& 0&0\\
0&-q& p&0\\
0& q&-p&0\\
0& 0& 0&0
\end{pmatrix}
_{\!\!n,n+1}, 
\quad
h_n=
\begin{pmatrix}
-\om_a& \om_d\\
 \om_a&-\om_d
\end{pmatrix}
_{\!\!n}
\end{align}
The subscripts at the bottom right of the matrix represent the vector space on which the matrix is acting. 
The Langmuir kinetics term $h_n$ can be regarded as an off-diagonal magnetic field, 
which induces a nonconservation of number of particles in the system.  

The stationary state $|S\ket$ belongs to the eigenvector of the Markov matrix with the eigenvalue zero:
\begin{align}
\mathcal{M}|S\ket=0. 
\end{align}
The first step is to find a stationary state $|S\ket$ explicitly. 
In order to do so, we introduce a basis transformation in the following. 

{\it Stationary state. }
Let $U$ be the $L$-fold tensor product of a $2\times2$ matrix defined by
\begin{align}
U=
\begin{pmatrix}
1& 1\\
\a&-1
\end{pmatrix}
^{\otimes L}
, \quad
\a=\frac{\om_a}{\om_d}, 
\end{align}
and consider the basis transformation 
\begin{align}
&\Mt=U^{-1}\mathcal{M}U
=\sum_{n=1}^{L}\(\Mt_{n,n+1}+\tilde{h}_n\), 
\nn\\&
\Mt_{n,n+1}=
\frac{1}{1+\a}
\nn\\&
\begin{pmatrix}
0& 0& 0&0\\
-\a(p-q)&-(q+p\a)& (p+q\a)&p-q\\
 \a(p-q)& (q+p\a)&-(p+q\a)&-(p-q)\\
0& 0& 0&0
\end{pmatrix}
_{\!\!n,n+1},
\nn\\&
\tilde{h}_n=
-\om
\begin{pmatrix}
0&0\\
0&1
\end{pmatrix}
_{\!\!n}, 
\end{align}
where we define $\om:=\om_a+\om_d$. 
It is noted that 
this transformation diagonalizes the Langmuir kinetics term. 
One of the most remarkable points in this transformation is that 
the stationary state in this basis becomes a simple pure vector $|00\cdots0\ket$. 
In the following we prove 
\begin{align}
\Mt|00\cdots0\ket=0.
\label{tm}
\end{align}
It is obvious that
\begin{align}
\tilde{h}_n|0\ket_{n}
=
-\om
\begin{pmatrix}
0&0\\
0&1
\end{pmatrix}
_{\!\!n}
\binom{1}{0}
_{\!\!n}
=0.
\end{align}
Moreover, we have
\begin{align}
\sum_{n=1}^{L}
\Mt_{n,n+1}|00\cdots0\ket
&=
\frac{\a(p-q)}{1+\a}
\sum_{n=1}^{L}
\(|n\ket-|n+1\ket\)
\nn\\&=0, 
\end{align}
where $|n\ket$ denotes the state with a single particle being at the site $n$. 
The last equality is due to the translational invariance of the system. This finishes the proof of \eqref{tm}. 

Thus the stationary state $|S\ket$ in the original basis is written as
\begin{align}
|S\ket&=
U|00\cdots0\ket
\nn\\&=
\binom{1}{\a}\otimes
\binom{1}{\a}\otimes
\cdots
\otimes\binom{1}{\a}
\nn\\&=
\sum_{N=0}^L
\a^N
\(
\sum_{\tau_1+\cdots+\tau_L=N}
|\tau_1,\cdots,\tau_L\ket
\). 
\end{align}
In the subspace with a fixed particle number $N$, 
the probability distributions of $\binom{L}{N}$ states 
are all proportional to $\a^N$. 
This means that $P(\tau_1,\cdots,\tau_L)$ is determined from particle number $N=\tau_1+\cdots+\tau_L$ only, 
independent of the explicit particle configuration $(\tau_1,\cdots,\tau_L)$. 

This result coincides with that given in \cite{EN} for the totally asymmetric case $q=0$. 
We note that our construction of the stationary state does not depend on the value of $q$. 

{\it Relaxation time. }
Next we construct the first excited state and evaluate the relaxation time. 
In the language of quantum spin chains, the stationary state in the transformed basis 
$|00\cdots0\ket$ is considered to be a completely ferromagnetic state. 
In this case the low lying excitations are given by spin wave states. 
According to this analogy, we can find low lying excitations of the ASEP-LK model. 

Let us define $S^\pm=\sg^x\pm\i\sg^y$, where $\sg^{x,y,z}$ denote standard Pauli matrices. 
These correspond to the ``spin flip operators." 
We further define total flip operators $S_T^\pm=\sum_{n=1}^LS_n^\pm$ and the state
\begin{align}
|S_T^-\ket&:=S_T^-|00\cdots0\ket
=\sum_{n=1}^L|n\ket, 
\end{align}
which is considered to be a spin wave excitation with the wave number $k=0$. 
This turns out to be the first excited state 
in the transformed basis. 
In the following we show that
\begin{align}
\Mt|S_T^-\ket=-\om|S_T^-\ket. 
\label{first}
\end{align}
We first note that $\Mt_{n,n+1}$ is written in the form 
\begin{align}
\Mt_{n,n+1}=
\begin{pmatrix}
0&0&0&0\\
-a&-b&c&d\\
a&b&-c&-d\\
0&0&0&0
\end{pmatrix}
_{\!\!n,n+1},
\label{mform}
\end{align}
where 
$a=\a(p-q)/(1+\a)$, 
$b=(q+p\a)/(1+\a)$, 
$c=(p+q\a)/(1+\a)$, 
and
$d=(p-q)/(1+\a)$. 
One can easily obtain
\begin{align}
\sum_{\ell=1}^L
&\Mt_{\ell,\ell+1}
|n\ket
=
a(|n,n+1\ket-|n-1,n\ket)
\nn\\&
+b(|n-1\ket-|n\ket)
+c(|n+1\ket-|n\ket), 
\end{align}
where 
$|n_1,n_2\ket$ denotes the state with two particles being at the sites $n_1$ and $n_2$:
$|n_1,n_2\ket:=S_{n_1}^-S_{n_2}^-|00\cdots0\ket$. 
This yields 
$\sum_{\ell=1}^L
\Mt_{\ell,\ell+1}
|S_T^-\ket=0$
from the translational symmetry. 
Together with the fact that 
$\sum_{\ell=1}^L
\tilde{h}_{\ell}
|S_T^-\ket=-\om|S_T^-\ket$, 
we can conclude the eigenvalue equation \eqref{first}. 

From direct diagonalizations of the Markov matrix for small systems, 
we observe that this eigenvalue $-\om$ is always the second largest real part among all the eigenvalues. 
Some examples are listed in Table \ref{exact_diag}. 
So we conclude that the relaxation time $T$ for the ASEP with Langmuir kinetics is given by
\begin{align}
T=\om^{-1}=\frac1{\om_a+\om_d}. 
\label{relax_time}
\end{align}
Note that this result is independent of system size $L$. 
Unlike the case with usual ASEP, the energy gap always has a finite value, 
which causes an exponentially fast relaxation to the stationary state. 

{\it Higher excitations. }
Next we construct further low lying excitations. 
In the following we show that 
\begin{align}
\Mt|(S_T^-)^m\ket=-m\om|(S_T^-)^m\ket 
\quad
(m=0, 1, 2, \cdots, L), 
\label{excite}
\end{align}
where
\begin{align}
|(S_T^-)^m\ket&:=\frac1{m!}(S_T^-)^m|00\cdots0\ket
\nn\\&=
\sum_{1\leq n_1<\cdots<n_m\leq L}|n_1,n_2,\cdots,n_m\ket. 
\end{align}
First we note that the following commutator 
\begin{align}
[\Mt_{n,n+1}, S_T^-]
=
\begin{pmatrix}
0&0&0&0\\
-a'&-b'&c'&d'\\
a'&b'&-c'&-d'\\
0&0&0&0
\end{pmatrix}
_{\!\!n,n+1}
\end{align}
has the same form as \eqref{mform}, 
where $a'=-(p-q)(1-\a)/(1+\a)$, $b'=-c'=-(p-q)/(1+\a)$, and $d'=0$. 
Since the derivation of $\sum_{\ell=1}^L
\Mt_{\ell,\ell+1}
|S_T^-\ket=0$
is independent of the explicit values of $a,b,c$ and $d$, 
we can inductively show that $\sum_{\ell=1}^L
\Mt_{\ell,\ell+1}
|(S_T^-)^m\ket=0$ with respect to $m$. 
Together again with 
$\sum_{\ell=1}^L
\tilde{h}_{\ell}
|(S_T^-)^m\ket=-m\om|(S_T^-)^m\ket$, 
we arrive at the eigenvalue equation \eqref{excite}. 

Direct diagonalizations for small systems show that, 
up to $(L+1)$ th largest eigenvalues, they are given by
$\lt0, -\om, -2\om, \cdots, -L\om \rt$ for sufficiently small $\om$. 
If we increase the value of $\om$, other complex eigenvalues $\lam$ with the real part $-\om>\text{Re} \, \lam>-L\om$ 
enter into this group. 
However, we observe that the second largest eigenvalue is always $-\om$. 
Some examples are listed in Table \ref{exact_diag}. 
\begin{table}
\begin{tabular}{|c|c|c|c|c|c|c|c|c|c|c|c|c|c|c|c|c|}
\hline
$\om$&\multicolumn{16}{c|}{Eigenvalues}\\
\hline
$0.2$ & $0$ & $0.2$ & $0.4$ &$0.6$ & $0.8$ & $z_1$ & $z_1^\ast$ &
$1.4$ & $1.4$ & $z_2$ & $z_2^\ast$ & $z_3$ & $z_3^\ast$ & $x_1$ & $x_2$ & $3.4$  
\\
\hline
$2$ & $0$ & $2$ & $z_4$ & $z_4^\ast$ & $4$ & $x_3$ & $5$ & $5$ & $x_4$ & $x_5$ & $6$ & 
 $z_5$ & $z_5^\ast$  & $7$ & $x_6$ & $8$ \\
\hline
\end{tabular}
\caption{
All the eigenvalues for $L=4$ with $\om=0.2$ and $2$. 
Numerical values in the table are given by 
\\
$z_1=1.27024 + 0.369916 \i$, 
$z_2=1.52976 + 0.369916 \i$, \\
$z_3=1.86397 + 0.287025 \i$, 
$x_1=2.31872$, 
$x_2=2.55334$, \\
$z_4=3.02539 + 0.243086 \i$, 
$x_3=4.10419$, 
$x_4=5.06102$, \\
$x_5=5.86836$, 
$z_5=6.97461 + 0.243086 \i$, 
$x_6=7.96644$. 
}
\label{exact_diag}
\end{table}

{\it Full dynamics. }
Finally let us consider the time evolution of physical quantities. 
Let $A$ be a physical quantity which takes a value $A=A(\tau)$ 
in a particle configuration 
$\tau:=(\tau_1, \cdots, \tau_L)$. 
We think of $A$ as an operator on a state vector $|\tau\ket$ 
acting diagonally as 
$A|\tau\ket=A(\tau)|\tau\ket$. 
Then the expectation value of the physical quantity $A$ in a state $|P\ket=\sum_\tau P(\tau)|\tau\ket$ is written as 
\begin{align}
\bra A \ket
&=
\frac{
\sum_\tau P(\tau)A(\tau)
}{
\sum_\tau P(\tau)
}
=
\bra T|A|P\ket
/
\bra T|P\ket, 
\end{align}
where $|T\ket:=\sum_\tau |\tau\ket$. 
Since the state vector $|P\ket$ obeys the master equation \eqref{master}, 
the time evolution of the physical quantity $A$ starting from the initial state $|I\ket$ reads
\begin{align}
\bra A(t) \ket
=
\bra T|Ae^{\mathcal{M}t}|I\ket
/
\bra T|I\ket. 
\end{align}
Here we have used $\bra T|e^{\mathcal{M}t}=\bra T|$, 
which follows from the property of the probability transition matrix 
$\bra T|M=0$. 

Hereafter we take the no-particle state $|00\cdots0\ket$ as an initial state $|I\ket$. 
Note that $\bra T|I\ket=1$. By the basis transformation the time evolution of $A$ can be written as 
\begin{align}
\bra A(t) \ket
=
\bra 00\cdots0|\At e^{\Mt t}|S\ket
\end{align}
where we have used 
$\bra \tilde{T}|=\bra T|U=(1+\a)^L\bra00\cdots0|$ 
and 
$|\tilde{I}\ket=U^{-1}|I\ket=(1+\a)^{-L}|S\ket$. 

We wish to expand the above quantity by the eigenvectors of Markov matrix $\mathcal{M}$. 
However, they are not necessarily orthogonal to each other 
since the Markov matrix $\mathcal{M}$ is not normal: 
$\mathcal{M}^\dagger\mathcal{M} \neq \mathcal{M}\mathcal{M}^\dagger$. 
However, we numerically confirm that the vectors $|(S_T^-)^m\ket$ are orthogonal to any other eigenvectors, 
and the rest of the eigenspace can always be orthogonalized by the Gram-Schmidt method. 
These facts lead to the following expansion:
\begin{align}
\bra A(t) \ket
&=
\sum_{m=0}^L
\frac{
\bra 00\cdots0|\At e^{\Mt t}|(S_T^-)^m\ket\bra(S_T^-)^m|S\ket
}{
\bra(S_T^-)^m|(S_T^-)^m\ket
}
\nn\\&
+\sum_{\lamt}
\frac{
\bra 00\cdots0|\At e^{\Mt t}|\lamt\ket\bra\lamt|S\ket
}{
\bra\lamt|\lamt\ket
}
. 
\end{align}
Noting that the stationary state $|S\ket$ can be expanded as
$|S\ket
=\sum_{m=0}^L
\a^m |(S_T^-)^m\ket 
$, 
we easily obtain
\begin{align}
\bra\lamt|S\ket=0 \quad \text{unless} \quad |\lamt\ket=|(S_T^-)^m\ket 
\quad 
(m=0,1,2,\cdots,L), 
\end{align}
where $|\lamt\ket$ are the eigenvectors of $\Mt$. 
Consequently we arrive at the formula for 
the time evolution of the physical quantity $A$ starting from the empty state $|00\cdots0\ket$ as follows: 
\begin{align}
\bra A(t) \ket
=\sum_{m=0}^L
e^{-m\om t}\a^m \bra 00\cdots0|\At|(S_T^-)^m\ket. 
\label{formula}
\end{align}

\begin{figure}
\includegraphics[width=0.9\columnwidth]{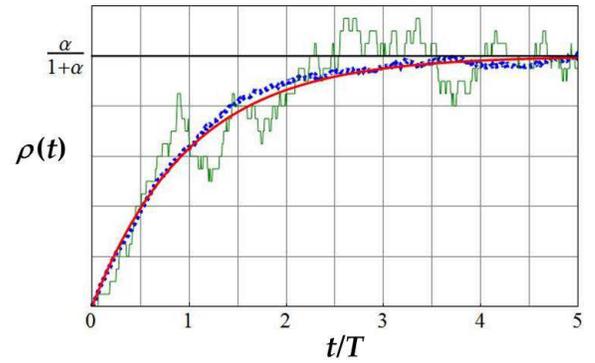}
\caption{
Time evolution of the particle density $\rho(t)$ starting from the empty state
is plotted against the scaled time $t/T$, where $T$ is the relaxation time \eqref{relax_time}. 
Parameters are chosen to be $(p,q)=(0.7,0.3)$, $(\om_a, \om_d)=(0.1,0.4)$. 
The thick red line is the plot of Eq. \eqref{density_formula}. 
The thin green line is the result of a one-time Monte Carlo simulation. 
The dashed blue line is a 100-times average of Monte Carlo simulations. 
We set the lattice length to $L=100$ in the simulation. 
The plot starts from zero and saturates to the value $\a/(1+\a)$ known as the Langmuir isotherm. 
}
\label{dens}
\end{figure}

\begin{figure}
\includegraphics[width=0.9\columnwidth]{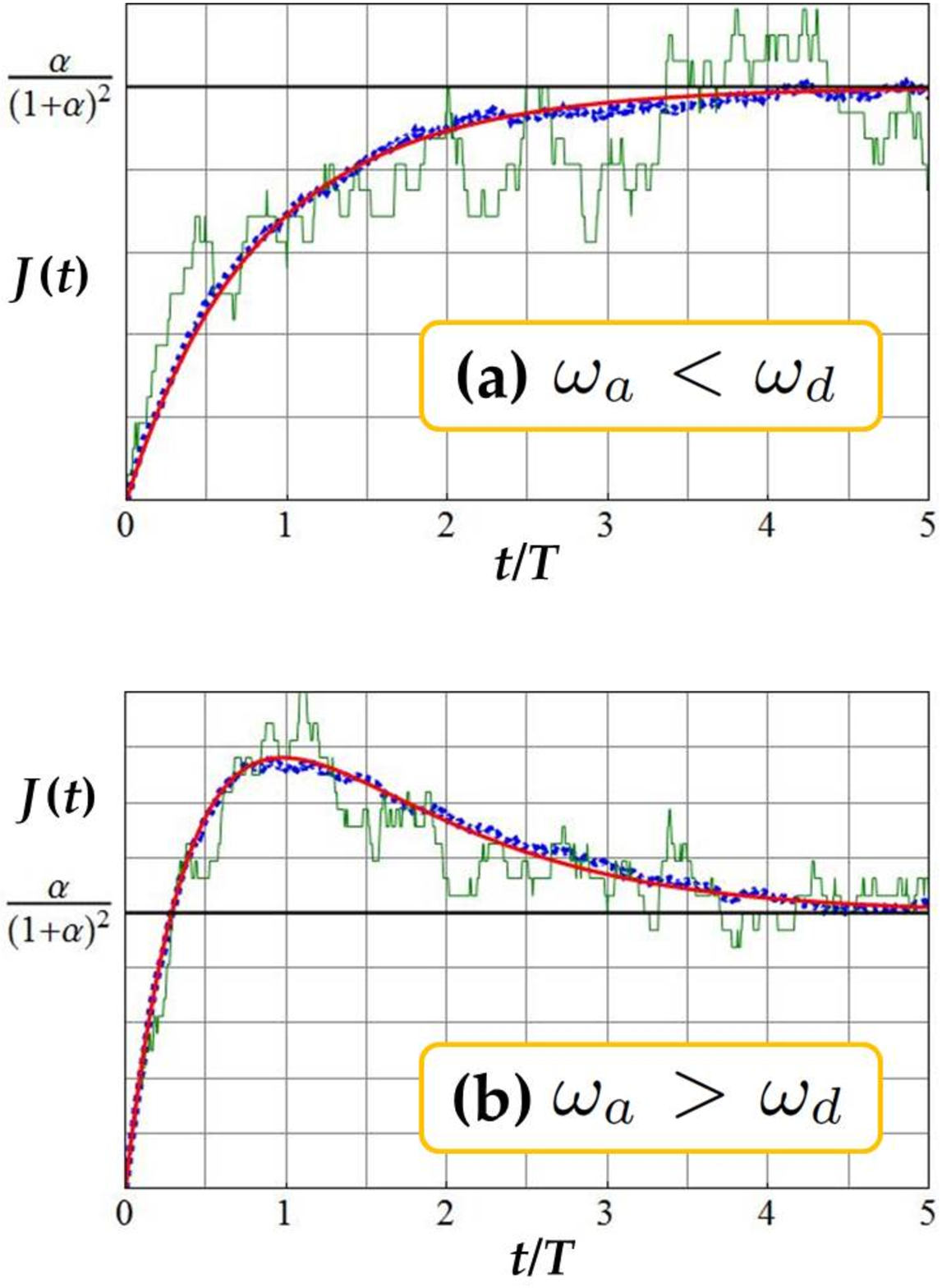}
\caption{
Time evolutions of the current $J(t)$ starting from the empty state
are plotted against the scaled time $t/T$ 
for (a) $\om_a<\om_d$ and (b) $\om_a>\om_d$. 
$T$ is the relaxation time \eqref{relax_time}. 
Parameters are chosen to be 
(a) $(p,q)=(0.7,0.3)$, $(\om_a, \om_d)=(0.1, 0.4)$ 
and 
(b) $(p,q)=(0.7,0.3)$, $(\om_a, \om_d)=(0.4, 0.1)$. 
The thick red line is the plot of Eq. \eqref{current_formula}. 
The thin green line is the result of a one-time Monte Carlo simulation. 
The dashed blue line is a 100-times average of Monte Carlo simulations. 
We set the lattice length to $L=100$ in the simulation. 
}
\label{crnt}
\end{figure}

{\it Density, current, and correlation functions. }
First let us apply the above formula to the particle density $\rho(t):=\bra \tau_1(t) \ket$,
 which represents the probability for a particle to exist at the site $1$. 
From the translational symmetry, 
multiplication of $L$ to this quantity 
yields the total number of particles in the system. 
Let $A$ be $A=(1-\sg_1^z)/2$. 
By the basis transformation we have 
$\tilde{A}=\dfrac1{1+\a}
\begin{pmatrix}
\a&-1\\-\a&1
\end{pmatrix}_{\!\!1}
$. 
Then we can easily obtain the time evolution of the particle density $\rho(t)=\bra A(t) \ket $ as 
\begin{align}
\rho(t)=\frac{\a}{1+\a}\(1-e^{-\om t}\), 
\label{density_formula}
\end{align}
which is shown in Fig. \ref{dens}, 
together with the Monte Carlo simulations. 
Note that this is an exact analytical result without any approximations. 
In the long time limit $t\to\infty$, it approaches the value $\a/(1+\a)$ 
known as the Langmuir isotherm \cite{Fowler}. 
Next let us consider the current $J(t):=\bra \tau_1(1-\tau_2)(t) \ket$ 
and set $A=(1-\sg_1^z)(1+\sg_2^z)/4$. 
By the basis transformation we have 
$\tilde{A}=\dfrac1{(1+\a)^2}
\begin{pmatrix}
\a&-1\\-\a&1
\end{pmatrix}_{\!\!1}
\begin{pmatrix}
1&1\\\a&\a
\end{pmatrix}_{\!\!2}
$. 
Substituting this into the Eq. \eqref{formula},  
we obtain the time evolution of the current, $J(t)=\bra A(t) \ket $, as 
\begin{align}
J(t)=\frac{\a}{(1+\a)^2}(1-e^{-\om t})(1+\a e^{-\om t}), 
\label{current_formula}
\end{align}
which is shown in Fig. \ref{crnt}, 
together with the Monte Carlo simulations. 
Note that $J(t)=\rho(t)(1-\rho(t))$, which means that the mean field approximation gives the exact dynamics. 
In the case $\om_a<\om_d$, the current monotonically increases and saturates to the value $\a/(1+\a)^2$. 
On the other hand, in the case $\om_a>\om_d$, the current is maximized to the value $1/4$ at $t=T\ln\frac{2\a}{\a-1}$, 
then begins to decrease and approaches the value $\a/(1+\a)^2$. 
This decrease of the current is due to the congestion of particles 
induced by the rapid increase of density by the large attachment rate $\om_a$. 
The steady state current $J(\infty)=\a/(1+\a)^2$ is maximized at the case $\a=1$, namely $\om_a=\om_d$. 

More generally, we can easily prove by use of Eq. \eqref{formula} 
that the time evolution of any correlation function 
$\bra x_1x_2\cdots x_n(t)\ket$ is decomposed into $\bra x_1(t)\ket\bra x_2(t)\ket\cdots \bra x_n(t)\ket$, 
where $x_\ell$ is $\tau_\ell$ or $1-\tau_\ell$. 
This decomposition results in 
\begin{align}
\bra x_1x_2\cdots x_n(t)\ket
=
\frac{\a^m}{(1+\a)^n}(1-e^{-\om t})^m(1+\a e^{-\om t})^{n-m}, 
\end{align}
where $m$ is the number of $\tau_\ell$'s in $\{x_1, x_2, \cdots, x_n\}$. 

{\it Conclusion. }
In this paper we examine the dynamical property of the ASEP with Langmuir kinetics on a periodic lattice. 
We introduce a basis transformation 
which enables us to construct the ground state and low lying excitations in a transparent way. 
As a result we obtain the exact stationary state and relaxation time. 
Furthermore, we analytically obtain the exact time evolution of the density and current 
starting from the initial state with no particle in the system. 
We also show that correlation functions with arbitrary length can be decomposed into the product of one-point functions. 
Relaxation dynamics from the other initial states and 
the transport property with an open boundary are 
the next problems to be solved. 
\acknowledgments
The authors thank S. Ichiki and K. Sakai for their useful discussions. 
This work was supported by JSPS KAKENHI Grant No. 25287026. 

\end{document}